%% file: main.tex
\documentclass[sigconf,screen]{acmart}

\copyrightyear{2026}
\acmYear{2026}
\setcopyright{cc}
\setcctype{by}
\acmConference[SIGIR '26]{Proceedings of the 49th International ACM SIGIR Conference on Research and Development in Information Retrieval}{July 20--24, 2026}{Melbourne, VIC, Australia}
\acmBooktitle{Proceedings of the 49th International ACM SIGIR Conference on Research and Development in Information Retrieval (SIGIR '26), July 20--24, 2026, Melbourne, VIC, Australia}
\acmDOI{10.1145/3805712.3808488}
\acmISBN{979-8-4007-2599-9/2026/07}

\usepackage{booktabs}
\usepackage{microtype}
\usepackage{amsmath}
\usepackage{url}

\usepackage[skip=0pt]{caption}
\usepackage[capitalise]{cleveref}
\usepackage[T1]{fontenc}

\usepackage{graphicx}
\usepackage{hyperref}
\usepackage{listings}
\usepackage{multirow}
\usepackage{subcaption}
\usepackage{tabularx}
\usepackage[dvipsnames]{xcolor}
\usepackage{colortbl}
\usepackage{calc}
\usepackage[inline]{enumitem}
\usepackage{numprint}
\usepackage[normalem]{ulem}

\newcommand{\qri}{\texttt{QRI}}
\newcommand{\bg}{\textsc{BG}}
\newcommand{\plain}{\textsc{P}}
\newcommand{\serp}{SERP}
\newcommand{\dcdata}{\texttt{HJM}}
\newcommand{\dcdatafullname}{Human-Judged Multilingual}
\newcommand{\logsdata}{\texttt{Logs}}

\setlist[itemize]{noitemsep,topsep=2pt,leftmargin=14pt}
\setlist[enumerate]{noitemsep,topsep=2pt,leftmargin=18pt}

\begin{document}

% \title[Behavior-Grounded LLM Judges for Search Evaluation]{Behavior-Grounded LLM Judges for Search Evaluation:\\ Auditable Telemetry Priors for Better Alignment with User Preferences}
\title[Aligning LLM Search Evaluation with Historical User Preferences]{As It Was: Aligning LLM Search Evaluation \\ with Historical User Preferences}

\author{Ali Vardasbi}
\affiliation{%
  \institution{Spotify}
  % \city{Amsterdam}
  \country{Netherlands}
}
\email{aliv@spotify.com}

\author{Gustavo Penha}
\affiliation{%
  \institution{Spotify}
  % \city{New York City}
  \country{United States}
}
\email{gustavop@spotify.com}

\author{Enrico Palumbo}
\affiliation{%
  \institution{Spotify}
  % \city{Turin}
  \country{Italy}
}
\email{enricop@spotify.com}

\author{Claudia Hauff}
\affiliation{%
  \institution{Spotify}
  % \city{Delft}
  \country{Netherlands}
}
\email{claudiah@spotify.com}

\author{Hugues Bouchard}
\affiliation{%
  \institution{Spotify}
  % \city{Barcelona}
  \country{Spain}
}
\email{hb@spotify.com}

\author{Mounia Lalmas}
\affiliation{%
  \institution{Spotify}
  % \city{London}
  \country{United Kingdom}
}
\email{mounia@acm.org}

\renewcommand{\shortauthors}{Ali Vardasbi et al.}
%% No italics, no superscripts, not anonymous
%% Use footnote or author note to identify equal contribution and/or contact author info

\begin{abstract}
\input{sections/00_abstract}
\end{abstract}




\maketitle

\input{sections/01_intro}
\input{sections/02_related}
\input{sections/03_method}
\input{sections/04_experiments}

\input{sections/05_results}

\input{sections/06_discussion}
\input{sections/ABtest}
\input{sections/07_conclusion}

\section{Presenter Bio}
Ali Vardasbi is a Research Scientist for Tech Research at Spotify, where he works on improving Search and Recommendation systems using LLMs. His work centers on evaluation, optimization, and advancing the effectiveness of LLM-powered experiences. 
% His research interests include agentic AI, efficiency in large-scale systems, and grounding techniques to make models more reliable and context-aware.

\section*{Acknowledgements}
The authors would like to thank Anders Nyman, Tyra Areskoug, Nicolo Felicioni, and Poppy Newdick for their invaluable assistance with data collection and for sharing their insights.

\balance
\bibliographystyle{ACM-Reference-Format}
\bibliography{bibliography}

\end{document}

%% file: sections/00_abstract.tex
Large-scale search systems evolve faster than human quality assurance scales,
especially for long-tail intents and multilingual queries.
LLM-as-a-judge approaches are a scalable alternative for evaluating the relevance of search engine result pages (\serp{}s), but judgments based solely on semantic similarity or world knowledge can drift from actual user preferences, particularly for ambiguous queries. 

We introduce a \emph{behavior-grounded} LLM judge that  augments each \serp{} item with a lightweight, auditable behavioral prior in the form of a \emph{Query--Relevance--Impressions} (\qri{}) card. 
Each card summarizes how users have historically interacted with similar queries and results, providing compact empirical evidence that the judge can cite to resolve ambiguity and make more consistent relevance judgments, while still relying on semantic reasoning.

In a large-scale music search evaluation at Spotify, using relevance estimates derived from historical user interactions across 6,000 recomposed \serp{}s, the behavior-grounded judge achieves stronger alignment with user preferences, improving Spearman rank correlation by approximately +5\% overall and yielding a +91\% relative improvement on disagreement cases.
On a multilingual human-judged dataset spanning five languages, grounding further increases correlation with human relevance judgments by +15\%. 
Importantly, when evaluated against outcomes from a live A/B test, the grounded judge shows consistently higher alignment with the observed winning model. 
While absolute alignment remains moderate, these findings demonstrate that lightweight behavioral grounding can improve the reliability and practical usefulness of LLM-based evaluation in real-world search systems.

%% file: sections/01_intro.tex
\section{Introduction}

%Search evaluation in production is constrained by two opposing forces: product surfaces and retrieval stacks change rapidly (new ranking features, new entity types, new generative interfaces), and at the same time, expert human labeling remains expensive and difficult to scale, especially for long-tail and multilingual queries. This has made \emph{LLM-as-a-judge} attractive as a scalable evaluation layer: given a query, user context, and a \serp{}, an LLM can produce a relevance assessment and a rationale.

Search systems in production evolve continuously, with frequent updates to ranking models, retrieval stacks, and user experiences. 
Ensuring that evaluation keeps pace with these changes requires scalable and reliable assessment methods.
As a result, \emph{LLM-as-a-judge} approaches have become attractive as a scalable evaluation layer, enabling relevance assessment and rationales directly from a query, context, and the \serp{}. This work is grounded in production music search at Spotify, where search surfaces, ranking models, and user expectations evolve rapidly across languages and regions.

\emph{Plain} LLM judges that rely solely on semantic similarity and catalog-centric reasoning may not fully capture how users interpret and engage with results~\cite{li2024llmsasjudges,jiang2025humanpreference}. %We refer to this setup as a \emph{plain} judge: a state-of-the-art LLM prompted with a standard relevance rubric but without access to historical interactions. 
While effective for clearly specified queries, their judgments can diverge from observed user preferences for underspecified or ambiguous queries, regionally dominant interpretations, and SERPs containing closely related variants.
%While strong on textual relevance and canonical correctness, these judges can be brittle for ambiguous or underspecified queries, regionally dominant interpretations, and SERPs composed of closely related variants. 
%In such cases, semantic correctness alone is often insufficient to reflect user preferences, and resulting judgments are difficult to audit.

Grounding through retrieval or external evidence has emerged as a common strategy for improving factual consistency and reliability in LLM systems~\cite{gao2023rag,ni2025trustworthy}.
Modern search systems already contain a rich signal of user intent in the form of aggregated interaction data, yet this signal is rarely incorporated into LLM-based evaluation. 
We argue that behavioral signals provide empirical evidence of how queries are interpreted in practice, including which results users engage with most frequently and how preferences are distributed across competing interpretations or variants.%\looseness=-1

To leverage this signal, we propose \textbf{behavior grounding} for LLM judges via aggregated interaction summaries. For each \serp{} item, we attach a \emph{Query--Relevance--Impressions} card that summarizes historically associated queries and their interaction statistics. 
The behavior-grounded judge is particularly useful when direct relevance estimates for the evaluated SERP are unavailable or unreliable. This occurs in long-tail or emerging queries with limited historical coverage, and in off-policy evaluation settings where the current ranking configuration has not been previously exposed to users.\looseness=-1
%The behavior-grounded judge is particularly useful in common production scenarios such as long-tail or emerging queries, and in off-policy evaluation, where direct relevance estimates for displayed results are unavailable or unreliable.
%The suggested use case of the behavior-grounded judge is evaluation in two common production scenarios: ({i}) novel or long-tail queries with insufficient interaction data, and ({ii}) off-policy evaluation where relevance estimates for displayed results are unavailable or invalid. In both cases, \qri{} cards provide partial, auditable evidence rather than a complete signal for the evaluated query.
% 
%Across our evaluation setups, behavior grounding consistently improves alignment with both user interaction signals and human judgments. 
%We also analyze when and why grounding changes evaluation outcomes, showing that its effects are most pronounced in cases where semantic-only reasoning is insufficient.

Across evaluation setups, behavior grounding improves alignment with both interaction signals and human judgments. Its impact is most pronounced when semantic reasoning alone is insufficient.

% \paragraph{Contributions.}
% \begin{itemize}
%   \item We introduce \qri{} cards: lightweight, auditable, per-result behavioral summaries for grounding LLM judges.
%   \item We describe a production-compatible pipeline to compute debiased query--entity relevance using IPS-style correction.
%   \item We present evidence that behavior grounding improves alignment with user preferences: a +5\% improvement in Spearman correlation with relevance estimated from historical user interactions in a large log-derived evaluation, with over a +91\% gain on cases where the two judges disagree, followed by an approximately +15\% increase in rank correlation with human relevance judgments in a multilingual evaluation.
%   \item We analyze \emph{when} grounding changes decisions, characterizing its effects on intent disambiguation, constraint handling, and sensitivity to behaviorally supported ranking differences.

% \end{itemize}

% \paragraph{Scope and privacy.}
% We study a real-world media search setting, but our approach is platform-agnostic:
% any product that has query logs and interaction telemetry can construct \qri{} cards.
% All behavioral features are aggregated over time windows and entity/query groups,
% and can be computed without exposing personal data.

%% file: sections/02_related.tex
\section{Related Work}

\paragraph{LLMs for evaluation.}
In Information Retrievl (IR), LLMs are increasingly used for scalable relevance labeling and scoring~\cite{li2024llmsasjudges,dli2025fromgeneration,vardasbi2026adaptive}.
Prior work highlights that LLM judgments can be poorly calibrated, prompt-sensitive, and may diverge from  human annotators in some settings~\cite{li2024llmsasjudges,jiang2025humanpreference}.
They may also hallucinate or drift when not grounded in external evidence~\cite{singh2025hallucinations,iwashima2025groundedresponses}.
Recent studies emphasize the importance of evidence-grounded and auditable evaluation mechanisms to improve reliability and alignment~\cite{genovese2026artificialauthority,sudjianto2024hcat}.
Our work builds on these insights by \emph{grounding} the LLM judge with compact, auditable behavioral evidence, with the goal of improving robustness and alignment with observed user preferences.

\paragraph{Behavioral signals and debiasing.}
User interaction logs provide rich implicit feedback (e.g., clicks), but are inherently biased~\cite{li2015counterfactual,zhuang2022implicit,vardasbi2020cascade}.
Counterfactual and propensity-based methods are widely used to correct these biases and obtain more reliable relevance estimates from logged data~\cite{zhu2020joint,saito2020unbiased,vardasbi2020when}.
However, even debiased estimates reflect historical interaction patterns rather than underlying semantic intent~\cite{wang2022causal}, which can limit their suitability as standalone relevance judgments. 
We therefore combine debiased query–entity relevance estimates with an LLM judge, using behavioral signals as grounded quantitative evidence.

\paragraph{Grounded decision-making and auditability.}
Grounding is a common strategy for improving reliability in LLM-based systems, often by incorporating external evidence through retrieval-augmented generation (RAG)~\cite{gao2023rag,ni2025trustworthy}.
Prior work highlights that grounding can improve factual consistency, transparency, and trustworthiness~\cite{gao2023rag,zhou2024trustworthiness}.
In industry evaluation workflows, auditability is critical: reviewers must be able to verify the evidence underlying a judgment~\cite{kentapadi2024grounding,ojewale2026audit}.
% Recent research further emphasizes attribution, provenance tracking, and audit trails as mechanisms for accountable LLM deployment~\cite{pang2025sourcing,hohensinner2026provenance}.
\qri{} cards are designed to support this requirement by providing interpretable behavioral summaries that fit within prompt budgets while remaining explicit enough to be cited in judge rationales, enabling grounded and reviewable evaluation decisions.

%% file: sections/03_method.tex
\section{Behavior-Grounded LLM Judge}

%We describe the behavior-grounded LLM judge and the behavioral evidence used to support its evaluations.

%\subsection{Problem Setup}

We formalize SERP evaluation as follows. Given a query $q$, optional context $c$ (e.g., locale),
and a \serp{} containing items $\mathcal{E}=\{e_1,\dots,e_n\}$, an LLM judge assigns a graded relevance label $y \in \{0, 0.5, 1\}$. 
Each result item is presented with standard descriptive metadata such as title, type, and basic attributes.

\paragraph{Evaluation variants.}
We compare two LLM-based evaluation variants that share the same relevance rubric, output space, and prompting structure, differing only in the information provided to the judge. The first, \textbf{Plain (P)}, evaluates the \serp{} using only the query, context, and item metadata, without access to behavioral evidence. Judgments are therefore based on semantic similarity, available metadata, and the model’s general knowledge. The second, \textbf{Behavior-Grounded (BG)}, receives the same inputs as \plain{}, augmented with a lightweight behavioral summary for each result item in the form of \qri{} cards. 
The prompt further specifies that \qri{} should be used as supporting behavioral context rather than as direct ground truth for the current query.

% These cards provide compact evidence of how users have interacted with similar queries and results, allowing the judge to ground relevance decisions in observed behavior while using the same rubric. \looseness=-1
%historically associated queries to $e_i$, debiased relevance estimates between the historical query and item, and impression volume.

%\subsection{\qri{} Cards: Query--Relevance--Impressions}
\paragraph{\qri{} Cards: Query--Relevance--Impressions}
QRI cards summarize historical interactions associated with each result item. They are constructed by aggregating a recent window of search logs, capturing how users engaged with similar queries and results prior to evaluation.
% Each log entry corresponds to a historical query $q'$ and an entity $e$ shown at a given rank position $p$. We aggregate both interacted and non-interacted impressions to record impression volume and downstream success signals (e.g., clicks or satisfying engagements). To account for position bias, we compute a debiased relevance estimate for each $(e, q', p)$ using inverse propensity scoring (IPS), following standard counterfactual estimation practices. This yields a relevance signal that reflects user preference while correcting for systematic exposure effects.
Each log entry corresponds to a historical query $q'\neq q$ and an entity $e \in \mathcal{E}$ shown in response. We aggregate both interacted and non-interacted impressions to record overall exposure volume and downstream engagement signals (e.g., clicks). From these data, we compute a debiased relevance estimate for each $(e, q')$ pair, yielding a signal that reflects user preference while mitigating systematic exposure effects.
Each $(e, q')$ pair contributes a single \qri{} line of the form $\{q' : \big(\hat{r}(e,q'),\; I(e,q')\big)\}$, where $\hat{r}$ is the relevance estimate and $I$ is the total impressions of $e$ for $q'$. This line is appended to the result item $e$ in the input of BG judge.

{
\paragraph{Debiased relevance estimate.}
We compute the debiased relevance estimate $\hat{r}(e, q')$ using an inverse propensity scoring (IPS) correction over result positions~\cite{joachims2017unbiased}. Propensities, representing the probability of examination at each rank, can be obtained from randomized experiments when available or estimated using standard click models. Our approach requires only a monotonic propensity curve and does not depend on a specific click model formulation.
}

% Our design aligns with prior work on grounding and attribution in LLM systems, which highlights the importance of transparent evidence sourcing and traceability~\cite{kentapadi2024grounding,pang2025sourcing}.

For a given result item, many historical queries may be associated with it. To limit prompt length, we rank these queries by semantic similarity to the evaluated query $q$ and retain only the top-$k$ entries. 
{
We set $k{=}10$ as a conservative cap to bound prompt length and prevent highly popular entities from dominating the evidence budget. In practice, many items have fewer than $k$ eligible historical queries after filtering, so the effective evidence size is often much smaller.
% As shown in \cref{fig:qri_count_alignment}, directional alignment benefits emerge already once the SERP has a small amount of behavioral support (for minimum thresholds $t\approx 5$--$9$, computed over the cumulative subset of SERPs with $\texttt{qri\_count}\ge t$, where $\texttt{qri\_count}$ is the total number of QRI lines summed across all entities on the page), suggesting the method is not reliant on large or densely populated \qri{} cards (and thus not on large effective evidence budgets per item).
}

\paragraph{Why \qri{} helps.}
% \qri{} cards provide an empirical prior that complements semantic reasoning by grounding relevance judgments in aggregated user behavior. This grounding offers context beyond surface similarity or global heuristics, indicating which entities users actually select when faced with comparable queries. As a result, \qri{} is particularly useful for stabilizing judgments in scenarios involving ambiguity, competing interpretations, and ranking trade-offs. We analyze these effects in detail in \cref{sec:discussion} based on observations from experiments on production data.
\qri{} cards provide an empirical prior that complements semantic reasoning by indicating which results users most frequently engage with for similar queries. 
%%SENTENCE BELOW IS ALSO IN INTRO
%This is most useful for ambiguous or underspecified queries and for cases where multiple plausible results exist and user behavior favors one over another (\cref{sec:discussion}).
Importantly, grounding does not override semantic or instructional reasoning. Instead, behavioral evidence is applied selectively: serving as a tie-breaker under ambiguity, or as a comparator among plausible candidates (\cref{sec:discussion}). For clearly specified queries, \qri{} typically plays a confirmatory rather than decisive role.

%% file: sections/04_experiments.tex
\section{Experimental Design}

We evaluate whether grounding LLM-based judges with behavioral evidence improves alignment with user preferences and human judgments, using real-world music search data from Spotify. 
All experiments are conducted on production SERPs, with behavioral summaries constructed from user interaction logs aggregated the month preceding evaluation.
Judgments are produced by a recent-generation commercial large language model accessed via API.

\subsection{Log-Derived Preference Evaluation}
To measure alignment with relevance estimated from historical user interactions at scale, we construct a log-derived evaluation dataset (hereafter, \logsdata{}). We sample approximately 5,000 queries issued by between 20 and 400 users over a 10-day period. This sampling ensures sufficient interaction volume for reliable relevance estimation while excluding highly popular head queries whose relevance is typically unambiguous. 

For each sampled query, we compute unbiased item-level relevance estimates from interaction logs. We then construct recomposed \serp{}s by recombining real items previously associated with each query, sampling 3--5 items per \serp{}. This recomposition creates \serp{}s with controlled differences in quality, including cases where highly relevant items (according to the log-derived estimates) are omitted or ranked lower than less relevant ones. 
The resulting dataset contains 5{,}965 \serp{}s spanning both high-quality and degraded configurations. 
%, as some queries are associated with multiple candidate \serp{}s.
% 
% \paragraph{Page-level relevance.}
Because the judge assigns labels at the page level, we compute a page relevance score
as a DCG-weighted average of item relevance estimates.
% followed by min--max normalization across the dataset.
% We evaluate whether pages assigned higher graded relevance by the judge are associated with higher interaction-derived relevance than those assigned lower grades.

\paragraph{Preventing query leakage.}
If a \qri{} card were to include the evaluation query itself, the BG judge could effectively recover prior user behavior, overstating the incremental contribution of grounding. 
%This would constitute a train--test leakage scenario in evaluation.
To prevent this, we exclude historical queries that are near-duplicates of the evaluation query, defined as having a cosine similarity greater than $0.9$.
% in the same embedding space used for top-$k$ selection. 
For example, for  the query \emph{``basketball warmup music''}, the near-duplicate \emph{``basketball warmup playlist''} is excluded, while \emph{``basketball training music''} is retained.
% similarly, for \emph{``calm morning''}, \emph{``calm happy morning''} is filtered out while \emph{``quiet morning''} remains eligible.
This filtering is applied only in the controlled evaluation setting to assess generalization. In production, near-duplicate queries are retained to leverage all available historical evidence.\looseness=-1
%to maximize alignment with user preferences.\looseness=-1

%this concern does not apply: the goal is not to estimate generalization performance, but to leverage all available historical evidence to maximize alignment with user preferences, making the inclusion of duplicate or near-duplicate queries both appropriate and desirable.\looseness=-1

% More examples:
% For \emph{lucid dreaming guided meditation}, \emph{lucid dreaming meditation} is
% excluded while \emph{lucid dreaming music} is retained.
% For \emph{music for focus}, \emph{music to focus} is excluded while
% \emph{background music for focus} is retained.
% For \emph{songs that make you want to dance}, \emph{songs that make you dance}
% is excluded while \emph{songs to dance to} is retained.
% For \emph{early 2000s rnb}, \emph{early 2000s r\&b} is excluded while
% \emph{early 2000s hip hop and r\&b} is retained.

% \paragraph{Evaluation.}
% We report Spearman's $\rho$ and Kendall's $\tau$
% between judge scores and interaction-derived page relevance.
% We further analyze the subset where the plain (\plain{}) and behavior-grounded (\bg{}) disagree, isolating the impact of grounding.

\subsection{Human-Judged Multilingual Evaluation}
Alignment with human annotators is a key evaluation criterion for LLM judges~\cite{jiang2025humanpreference,sudjianto2024hcat}.
We therefore assess whether behavior grounding also improves agreement with human relevance judgments, using an internal multilingual dataset of 265 \serp{} instances across five languages, referred to as the \dcdata{} (\dcdatafullname) dataset.
Each instance is annotated by human raters using a three-level graded relevance scale.\\
% For consistency with the rest of our analysis, we normalize these labels to numeric values in $\{0, 0.5, 1\}$.

For both \logsdata{} and \dcdata{}, we report Spearman's $\rho$ and Kendall's $\tau$ between judge predictions and human-assigned relevance scores, and analyze the subset of instances where \plain{} and \bg{} disagree.

%We evaluate against an internal multilingual dataset consisting of 265 \serp{} instances across five languages, which we refer to as the \dcdata{} (\dcdatafullname) dataset.
%Each instance is annotated by human raters using a three-level graded relevance scale.
%For consistency with the rest of our analysis, we normalize these labels to numeric values in $\{0, 0.5, 1\}$.
%We report both Spearman's $\rho$ and Kendall's $\tau$, and analyze the subset where \plain{} and \bg{} disagree. 

% Unlike the log-derived setting, human judgments are provided on a discrete three-level relevance scale rather than a continuous page-relevance score.
% ; accordingly, the output domain of the LLM judges and the human annotations is identical in this evaluation.

%% file: sections/05_results.tex
%\section{Results}

% We evaluate the behavior-grounded (BG) judge along three complementary axes:
% (i) alignment with historical user interactions (\logsdata{}),
% (ii) agreement with curated human judgments (\dcdata{}), and
% (iii) consistency with outcomes from an online A/B test.

\input{resources/tab_corr-all}

% \subsection{Alignment with Logs-Derived Relevance}
%\subsection{Alignment with Logs-Derived Relevance and Human Judgments}
\section{Alignment with Logs-Derived Relevance and Human Judgments}
We first evaluate alignment with historical user preferences using the \logsdata{} dataset.
On the full set of instances, the behavior-grounded (BG) judge achieves higher correlation with page relevance than the plain (P) judge (Table~\ref{tab:corr-all}, \logsdata{}, All),
indicating improved alignment with user preference at scale.
% \add{
% While the overall correlation gains are modest (e.g., $\rho$: 0.416$\rightarrow$0.438 on \logsdata{}),
% they concentrate on the decision-critical subset where the judge changes its label (Flipped), where BG more than doubles rank correlation.
% In practice, these are the cases most likely to affect model comparisons and ranking diagnostics.
% Consistently, BG also improves online sign alignment by +6.2 points absolute (30.6\%\(\rightarrow\)36.8\%), indicating a measurable reduction in directional evaluation errors even when absolute predictability remains challenging.
% }
% 
The effect of behavior grounding is most pronounced on flipped instances (\logsdata{}, Flipped),
where the two judges assign different graded relevance levels.
On this subset, BG substantially outperforms P, achieving more than a twofold increase in both Spearman’s $\rho$ and Kendall’s $\tau$. This suggests that grounding is particularly effective in cases where semantic-only  reasoning is insufficient to resolve ambiguity or ranking trade-offs. In contrast, on the much larger subset of instances where the two judges agree (\logsdata{}, Equal),
both variants exhibit similarly strong correlation with interaction-derived relevance.
This indicates that grounding concentrates its impact on ambiguous or contested cases without degrading performance on  straightforward evaluations.

\input{resources/fig_results}

% \subsection{Alignment with Human Judgments}

We next assess alignment with human judgments using the \dcdata{} dataset. 
As in the log-derived evaluation, BG achieves higher rank correlation with human-assigned graded relevance than the P judge (Table~\ref{tab:corr-all}, \dcdata{}, All), indicating more consistent ordering of pages by perceived usefulness.
Absolute correlations remain moderate, leaving a substantial gap to perfect agreement with human annotators.
% Behavior grounding therefore improves alignment, but does not fully close the gap between automated and human evaluation.
% 
% 
%Table~\ref{tab:corr-all} (\dcdata{}, All) shows that BG improves both Spearman’s $\rho$ and Kendall’s $\tau$ over P.
%This improvement reflects better calibration of graded relevance levels rather than gains confined to a single label.
%When restricting attention to instances where the two judges produce identical predictions (\dcdata{}, Equal),
%both judges exhibit strong correlation with human judgments.
%However, on the small but decision-critical subset where P and BG disagree (\dcdata{}, Flipped),
%BG shows substantially stronger correlation, while P exhibits near-zero or negative correlation.
%Again, when grounding changes a decision, it tends to move the prediction toward human judgment.
% 
However, on the disagreement subset (\dcdata{}, Flipped), BG exhibits substantially higher correlation with human judgments, while P shows negative correlation. 
As in the log-derived setting, grounding consistently shifts predictions closer to human judgments when it changes a decision.

\paragraph{Diagnostic Analysis of Flipped Instances} 
Figure~\ref{fig:bucket_comparison} analyzes flipped instances (P $\neq$ BG), where grounding alters the predicted label. 
Figure~\ref{fig:logs_violin} shows the distribution of interaction-derived page relevance conditioned on the graded relevance level predicted by each judge. 
For pages assigned the highest level (label=1), BG’s predictions are consistently associated with higher relevance than those of P. 
Conversely, pages that BG assigns low relevance labels tend to have lower underlying page relevance than pages receiving the same low labels from P. 
% These patterns indicate that, on flipped cases, BG’s relevance assignments better reflect observed user preferences.
%Inspecting the mean values (shown as dashed lines), we further observe that instances labeled as non-relevant (label=0) or partially relevant (label=0.5) by BG have lower mean page relevance than those assigned the same levels by P, while for the highest relevance level (label=1) the opposite holds.
%Together, these patterns indicate that BG’s predictions on flipped cases are more closely aligned with observed user preferences.
% \input{resources/fig_logs_buckets}
% \input{resources/fig_sb_buckets}

Figures~\ref{fig:logs_buckets} and~\ref{fig:sb_buckets} further decompose this effect across page-relevance quantiles.
Across both datasets, BG assigns high relevance more frequently to pages in the top relevance buckets
and avoids over-assigning high relevance to pages with low interaction-derived or human-assigned relevance. 
% The same qualitative pattern appears in both the large log-derived dataset and the smaller curated dataset, despite their different construction and scale.

Taken together, these diagnostics indicate that grounding reallocates labels toward empirically supported pages, particularly  where semantic-only judgments are least reliable.
% This finding is consistent with prior evidence that grounding improves factual consistency and trustworthiness in LLM-based systems~\cite{gao2023rag,ni2025trustworthy}.

%% file: resources/tab_corr-all.tex
\begin{table}[t]
  \centering
  \footnotesize
  \caption{Correlation between judge predictions and relevance.
  ``Flipped'' refers to instances where P and BG disagree; ``Equal'' denotes agreement.}
  \label{tab:corr-all}
  \begin{tabular}{llrcccc}
    \toprule
    \multirow{2}{*}{Dataset} & \multirow{2}{*}{Subset} & \multirow{2}{*}{Size}
    & \multicolumn{2}{c}{Spearman $\rho$} & \multicolumn{2}{c}{Kendall $\tau$} \\
    \cmidrule(lr){4-5} \cmidrule(lr){6-7}
    & & & P & BG & P & BG \\

    \midrule
    Logs & All     & 5965 & 0.416 & \textbf{0.438} & 0.336 & \textbf{0.354} \\
    Logs & Flipped & 918  & 0.147 & \textbf{0.281}$^\dagger$ & 0.114 & \textbf{0.221}$^\dagger$ \\
    Logs & Equal   & 5047 & \multicolumn{2}{c}{0.457} & \multicolumn{2}{c}{0.371} \\
    \midrule
    \dcdata{} & All     & 265  & 0.450 & \textbf{0.516} & 0.413 & \textbf{0.476} \\
    \dcdata{} & Flipped & 27   & -0.127 & \textbf{0.530}$^\dagger$ & -0.108 & \textbf{0.486}$^\dagger$ \\
    \dcdata{} & Equal   & 238  & \multicolumn{2}{c}{0.525} & \multicolumn{2}{c}{0.486} \\
    \bottomrule
  \end{tabular}
  \\
\footnotesize $\dagger$: statistically significant improvement ($p < 0.05$
\vspace{-1.5em}
).
\end{table}

%% file: resources/fig_results.tex
% 0.3043137255
% 0.3803921569
% 0.2852941176

\begin{figure*}[t]
  \centering

  \begin{subfigure}[t]{0.305\textwidth}
    \centering
    \includegraphics[scale=0.27]{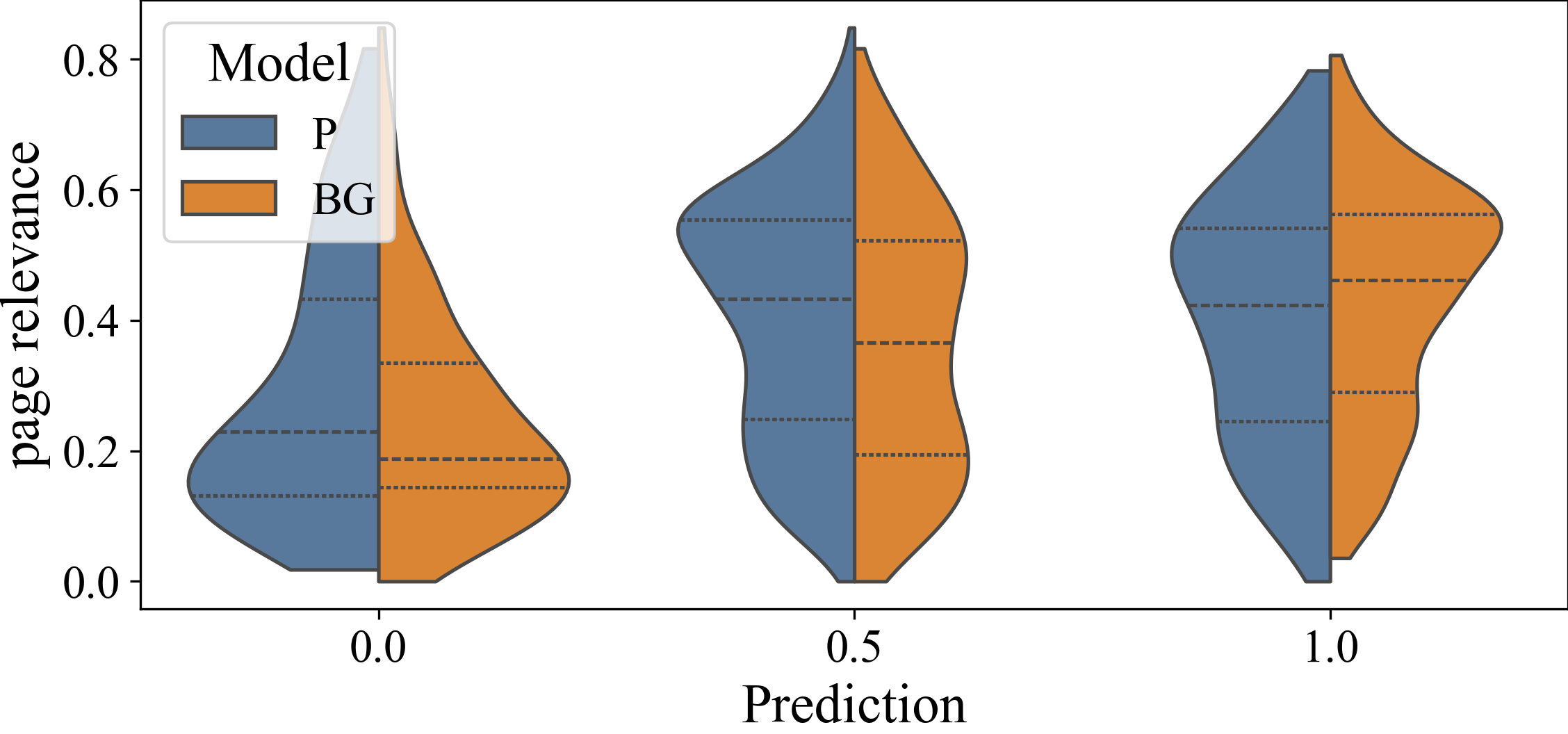}
    \vspace{-0.6em}
    \caption{Flipped \logsdata{}: interaction-derived relevance vs.\ judge label.}
    \label{fig:logs_violin}
  \end{subfigure}
  \hfill
  \begin{subfigure}[t]{0.382\textwidth}
    \centering
    \includegraphics[scale=0.27]{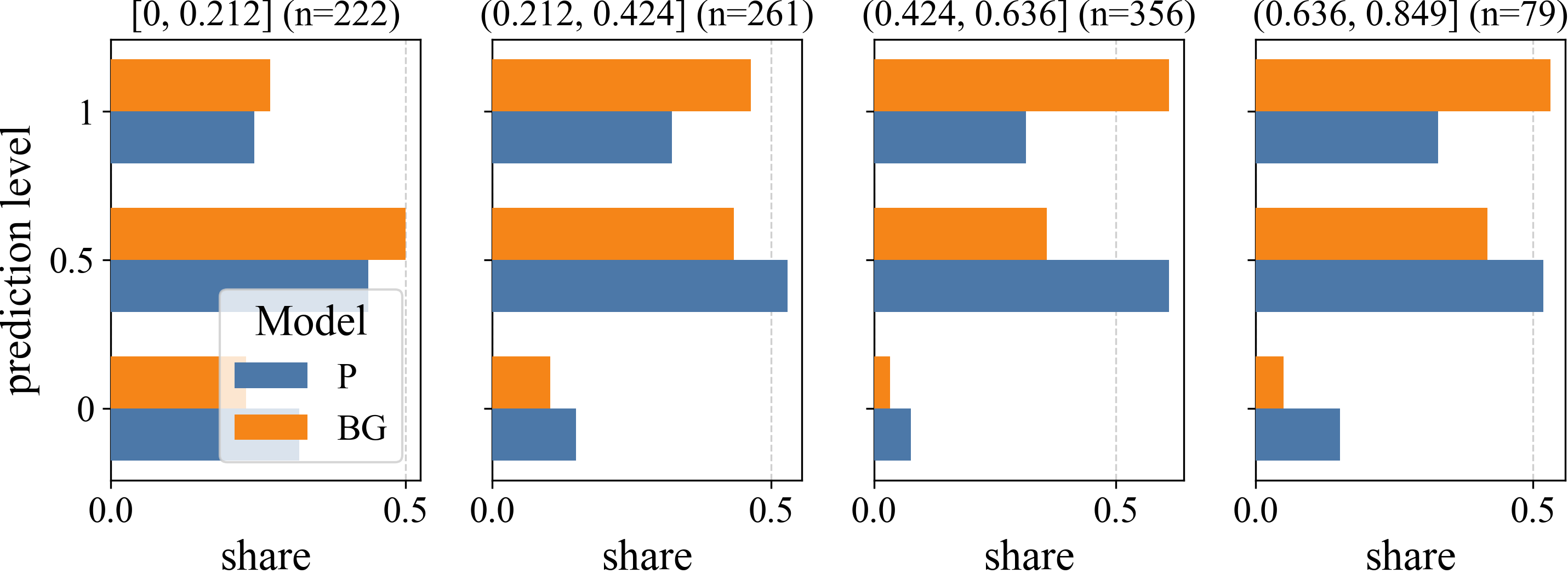}
    \vspace{-0.6em}
    \caption{Flipped \logsdata{}: relevance buckets.}
    \label{fig:logs_buckets}
  \end{subfigure}
  \hfill
  \begin{subfigure}[t]{0.287\textwidth}
    \centering
    \includegraphics[scale=0.27]{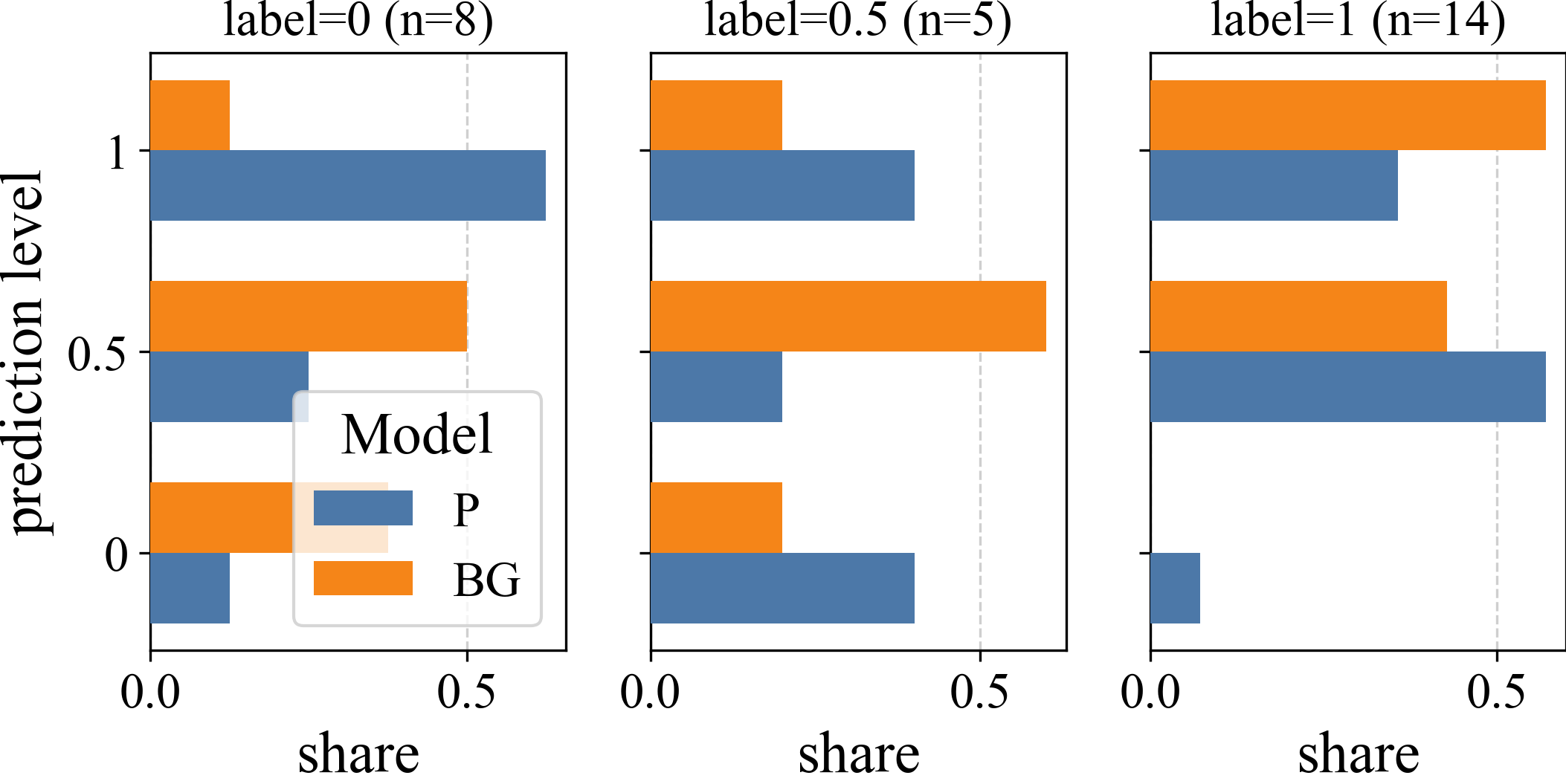}
    \vspace{-0.6em}
    \caption{Flipped \dcdata{}: human labels.}
    \label{fig:sb_buckets}
  \end{subfigure}
  
  \vspace{-0.3em}
  \caption{Diagnostics on flipped instances (where P and BG disagree).}
  \label{fig:bucket_comparison}
\end{figure*}

%% file: sections/06_discussion.tex
\section{Analysis of Grounding Effects}
\label{sec:discussion}

% All analyses and experiments in this section are conducted on real-world music search data from Spotify.
% including production search SERPs and associated user interaction logs. 
% To ensure coverage across varying quality levels, the evaluation set includes perturbed SERPs in which top-ranked results were intentionally removed, enabling analysis of both high-quality and degraded ranking scenarios.

Our findings show that \qri{} grounding changes judgments in systematic and interpretable ways, rather than inducing a uniform shift toward higher or lower relevance labels. Across both datasets, BG diverges from P along recurring patterns: resolving ambiguous intent, recalibrating the severity of near-miss cases, and adjusting judgments when multiple plausible results are ranked differently. 
These differences are auditable, as they can be traced directly to specific \qri{} evidence surfaced to the judge.

%\subsection{Ambiguity and Severity Calibration}
\paragraph{Resolving ambiguity.} 
% \subsection{Where Behavior Helps Most: Ambiguity and Severity Calibration}
Grounding is most impactful for ambiguous or underspecified queries, such as lyric fragments, short titles with multiple plausible referents, or regionally overloaded terms. In these cases, \qri{} provides of how users interpret the query in practice. BG often anchors a different primary entity than P, based on on observed engagement patterns. 
For example, for the query \emph{``when you say you love me''}, P assumes a title-based search and penalizes the SERP for missing a track with that exact title. In contrast, \qri{} shows strong engagement with Miley Cyrus’ \emph{Adore You},  suggesting that users interpret the query as a lyric search. BG therefore assigns a higher relevance score to the SERP accordingly.

\paragraph{Calibrating severity.} Grounding also affects how strongly errors are penalized in near-miss scenarios. When \qri{} indicates strong demand for a specific entity, BG is stricter than P if that entity is missing or poorly ranked. 
For example, for query \emph{``dark til daylight''}, P assigns partial credit because thematically related results are present, whereas BG assigns a clear failure after observing that the track with the exact same title (by Morgan Wallen) is absent and substitutes receive little engagement from similar historical queries. 
Conversely, when \qri{} indicates that users consistently engage with related but indirect results, BG may assign a more lenient judgment.
For instance, if a playlist surfaces the intended track prominently,
BG treats the SERP as acceptable in practice rather than than applying a stricter penalty.\looseness=-1

%Together, these cases show that behavioral evidence is most valuable not for obvious queries, but for resolving ambiguity and calibrating severity in near-miss scenarios, capabilities that are difficult to achieve using semantic rules alone.

%\subsection{Ranking Sensitivity}
% \subsection{Behavior as a Tie-Breaker: Ranking Sensitivity}

\paragraph{Ranking sensitivity.} A third effect of grounding is increased sensitivity to ranking quality. When multiple plausible results are present, \bg{} compares their relative \qri{} strength and penalizes cases where the entity most strongly favored by user engagement is not ranked first, even when \plain{} considers the SERP acceptable.
For example, if a later re-recording is ranked above the historically preferred original,
% (e.g., \emph{``99 Luftballons''}), 
BG is more likely to assign a lower relevance score to the page by recognizing that user interactions consistently favor the original version. In such cases, grounding reinforces ranking distinctions that are weakly signaled by semantics alone but strongly supported by user behavior.

%% file: sections/ABtest.tex
\section{Online A/B Test Alignment}

Beyond offline alignment with interaction and human signals, we evaluate whether grounding improves agreement with live system outcomes. We therefore test the judges against results from a production A/B experiment comparing two ranking systems (Model A and Model B). The experiment ran for one week, from which we sampled 904 queries.

Behavioral evidence used to construct \qri{} cards was drawn from one month of historical search interactions ending one week prior to the A/B test, ensuring no temporal overlap between interaction signals and evaluation outcomes. As in our offline setup, we applied the same leakage-avoidance procedure (cosine similarity threshold of $0.9$) when constructing \qri{} cards. 
For each query, we sampled three \serp{} instances from each model. Judge predictions and online outcomes were aggregated at the query level by averaging across the three instances per model.
We then measure \emph{sign alignment}: whether the judge predicts the same winning model per query as observed online. As shown in Table~\ref{tab:ab_sign_alignment}, BG achieves higher alignment than P,
% and overall, 
and the difference is statistically significant. 
% Nevertheless, alignment remains modest for both judges, indicating that offline evaluation remains an imperfect predictor of online impact.

\begin{table}[t]
\small
\centering
\caption{Sign alignment with online A/B outcomes.}
\label{tab:ab_sign_alignment}
\begin{tabular}{ccc}
\toprule
Query Count & P aligned (\%) & BG aligned (\%) \\
\midrule
% Segment 1 & 31.8 & \textbf{37.3$^\dagger$} \\
% Segment 2 & 28.7 & \textbf{36.0$^\dagger$} \\
904   & 30.6 & \textbf{36.8$^\dagger$} \\
\bottomrule
\end{tabular}
\\
\footnotesize $\dagger$: statistically significant improvement ($p < 0.01$).

\end{table}

\input{resources/qri_count}

%\paragraph{Effect of Behavioral Signal Strength.}

To understand when behavior grounding is most beneficial, we further analyze performance as a function of $\texttt{qri\_count}$, the total number of historical queries included across all \qri{} cards in a \serp{}. 
This measure reflects the amount of available behavioral support for the evaluated items—a loose proxy for their historical exposure and relative popularity.

Online results indicate that Model~A outperforms Model~B (i.e., $\Delta(y)>0$). 
Figure~\ref{fig:qri_count_alignment} illustrates the evolution of the judges’ implied preference difference, $\Delta(\hat{y})$, as a function of the $\texttt{qri\_count}$ threshold.
Specifically, each point on the $x$-axis represents a cumulative subset; for instance, the value at $x=5$ aggregates all queries with at least five QRI cards.
At low $\texttt{qri\_count}$, BG remains closer to the neutral (zero) line, indicating smaller directional error, whereas P exhibits a larger deviation. As behavioral support increases, BG crosses the zero threshold earlier and moves more decisively into the positive region, aligning with the online result at moderate levels of support. 
Specifically, BG becomes directionally consistent with the online preference for thresholds $\texttt{qri\_count} \ge 5$, with statistically significant positive alignment by $\texttt{qri\_count} \ge 9$. 
In contrast, P approaches zero more gradually and does not reach statistically significant positive alignment.
{
Low $\texttt{qri\_count}$ corresponds to the cold-start/long-tail regime where behavioral evidence is sparse; here BG largely falls back to semantic reasoning.
At the other extreme, the highest $\texttt{qri\_count}$ bins contain relatively few queries (see the count bars), so the estimated $\Delta(\hat y)$ is higher-variance and more sensitive to bucket composition.
}
%Taken together, these findings indicate that grounding improves alignment not only with interaction data and human judgments, but also with A/B outcomes. At the same time, absolute alignment rates remain moderate, leaving a substantial gap to oracle-level evaluation.
Overall, these results suggest that behavior grounding enables faster and more reliable convergence toward the observed online preference as behavioral evidence accumulates.\\

Taken together, these findings indicate that grounding improves alignment not only with interaction data and human judgments, but also with live A/B outcomes. Absolute alignment remains moderate, reflecting the difficulty of predicting online preferences. Further gains may require richer behavioral signals.

%% file: resources/qri_count.tex
\begin{figure}[t]
  \centering
  \includegraphics[width=0.9\linewidth]{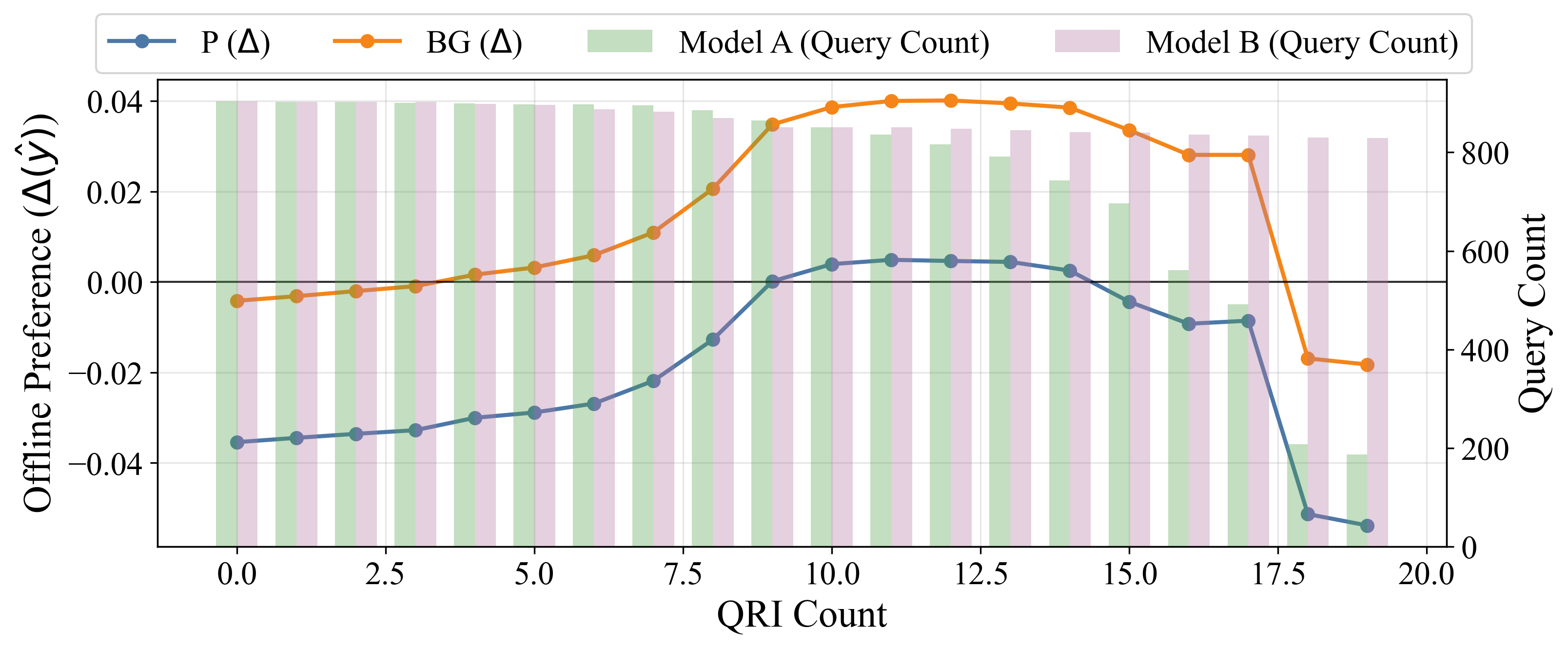}
  \caption{
  Offline judge preference $\Delta(\hat{y})$ as a function of the minimum SERP-level $\texttt{qri\_count}$ threshold $t$. Each data point represents a cumulative subset of queries where $\texttt{qri\_count} \ge t$. The online ground truth indicates a preference for Model A ($\Delta(y) > 0$).
  }
  \vspace{-0.5em}
  \label{fig:qri_count_alignment}
\end{figure}

%% file: sections/07_conclusion.tex
\section{Conclusion}

We studied the impact of incorporating user behavior signals into LLM-based evaluation of music search SERPs at Spotify. 
Our framework bridges counterfactual estimation from interaction logs~\cite{li2015counterfactual} with grounding mechanisms developed for reliable LLM systems~\cite{gao2023rag}.
By comparing a plain semantic judge with a behavior-grounded variant multiple evaluation settings—including interaction-derived relevance, human judgments, and live A/B outcomes, we showed that behavioral evidence leads to systematic and interpretable changes in evaluation decisions, rather than indiscriminate shifts toward higher or lower relevance labels.

Behavior grounding is most impactful in three recurring scenarios: resolving ambiguous intent, calibrating the severity of near-miss cases, and increasing sensitivity to ranking differences among plausible results. These findings highlight the value of behavior-grounded LLM evaluation in practice. Behavioral signals complement rubric-driven judgment by providing targeted, auditable evidence that improves alignment with observed user preferences.

%We found behavior grounding to be most impactful in three scenarios: resolving ambiguous intent, calibrating the severity of near-miss cases, and increasing sensitivity to ranking differences among plausible results.  %In these scenarios, interaction signals help identify the entities users actually seek, distinguish serious failures from tolerable imperfections, and reinforce ranking distinctions that semantic reasoning alone may underspecify. 

%These findings highlight both the promise and the boundaries of behavior-grounded LLM evaluation. Behavioral signals do not replace rubric-driven judgment; instead, they serve as targeted, auditable evidence that improves alignment with real user preferences. Future work should explore stronger guardrails to ensure behavioral evidence is used appropriately and does not amplify residual biases in interaction logs, as well as improved methods for correcting exposure effects and extending this grounding approach to other retrieval domains and evaluation tasks.

Future work should investigate stronger safeguards to ensure behavioral evidence is used appropriately and does not amplify residual biases in interaction logs, as well as improved debiasing and aggregation methods. 

%% file: bibliography.bib
@inproceedings{joachims2017unbiased,
author = {Joachims, Thorsten and Swaminathan, Adith and Schnabel, Tobias},
title = {Unbiased Learning-to-Rank with Biased Feedback},
year = {2017},
isbn = {9781450346757},
publisher = {Association for Computing Machinery},
address = {New York, NY, USA},
url = {https://doi.org/10.1145/3018661.3018699},
doi = {10.1145/3018661.3018699},
booktitle = {Proceedings of the Tenth ACM International Conference on Web Search and Data Mining},
pages = {781–789},
numpages = {9},
location = {Cambridge, United Kingdom},
series = {WSDM '17}
}

@misc{li2024llmsasjudges,
      title={LLMs-as-Judges: A Comprehensive Survey on LLM-based Evaluation Methods}, 
      author={Haitao Li and Qian Dong and Junjie Chen and Huixue Su and Yujia Zhou and Qingyao Ai and Ziyi Ye and Yiqun Liu},
      year={2024},
      eprint={2412.05579},
      archivePrefix={arXiv},
      primaryClass={cs.CL},
      url={https://arxiv.org/abs/2412.05579}, 
}

@inproceedings{dli2025fromgeneration,
    title = "From Generation to Judgment: Opportunities and Challenges of {LLM}-as-a-judge",
    author = "Li, Dawei  and
      Jiang, Bohan  and
      Huang, Liangjie  and
      Beigi, Alimohammad  and
      Zhao, Chengshuai  and
      Tan, Zhen  and
      Bhattacharjee, Amrita  and
      Jiang, Yuxuan  and
      Chen, Canyu  and
      Wu, Tianhao  and
      Shu, Kai  and
      Cheng, Lu  and
      Liu, Huan",
    editor = "Christodoulopoulos, Christos  and
      Chakraborty, Tanmoy  and
      Rose, Carolyn  and
      Peng, Violet",
    booktitle = "Proceedings of the 2025 Conference on Empirical Methods in Natural Language Processing",
    month = nov,
    year = "2025",
    address = "Suzhou, China",
    publisher = "Association for Computational Linguistics",
    url = "https://aclanthology.org/2025.emnlp-main.138/",
    doi = "10.18653/v1/2025.emnlp-main.138",
    pages = "2757--2791",
    ISBN = "979-8-89176-332-6",
}

@article{jiang2025humanpreference,
    author = {Jiang, Ruili and Chen, Kehai and Bai, Xuefeng and He, Zhixuan and Li, Juntao and Yang, Muyun and Zhao, Tiejun and Nie, Liqiang and Zhang, Min},
    title = {A Survey on Human Preference Learning for Aligning Large Language Models},
    year = {2025},
    issue_date = {April 2026},
    publisher = {Association for Computing Machinery},
    address = {New York, NY, USA},
    volume = {58},
    number = {6},
    issn = {0360-0300},
    url = {https://doi.org/10.1145/3773279},
    doi = {10.1145/3773279},
    journal = {ACM Comput. Surv.},
    month = dec,
    articleno = {152},
    numpages = {39}
}

@INPROCEEDINGS{singh2025hallucinations,
    author={Singh, Ravneet and Singh, Parminder and Malik, Arun and Sukmawan, Dede},
    booktitle={2025 International Conference on Metaverse and Current Trends in Computing (ICMCTC)}, 
    title={Understanding and Mitigating Hallucinations in Large Language Models: Insights from a Systematic Literature Review}, 
    year={2025},
    volume={},
    number={},
    pages={1-10},
    doi={10.1109/ICMCTC62214.2025.11196493}
}

@misc{iwashima2025groundedresponses,
      title={Factors That Support Grounded Responses in LLM Conversations: A Rapid Review}, 
      author={Gabriele Cesar Iwashima and Claudia Susie Rodrigues and Claudio Dipolitto and Geraldo Xexéo},
      year={2025},
      eprint={2511.21762},
      archivePrefix={arXiv},
      primaryClass={cs.CL},
      url={https://arxiv.org/abs/2511.21762}, 
}

@Article{genovese2026artificialauthority,
    AUTHOR = {Genovese, Ariana and Hegstrom, Lars and Prabha, Srinivasagam and Gomez-Cabello, Cesar A. and Haider, Syed Ali and Collaco, Bernardo and Wood, Nadia G. and Forte, Antonio Jorge},
    TITLE = {Artificial Authority: The Promise and Perils of LLM Judges in Healthcare},
    JOURNAL = {Bioengineering},
    VOLUME = {13},
    YEAR = {2026},
    NUMBER = {1},
    ARTICLE-NUMBER = {108},
    URL = {https://www.mdpi.com/2306-5354/13/1/108},
    PubMedID = {41596039},
    ISSN = {2306-5354},
    DOI = {10.3390/bioengineering13010108}
}

@misc{sudjianto2024hcat,
      title={Human-Calibrated Automated Testing and Validation of Generative Language Models}, 
      author={Agus Sudjianto and Aijun Zhang and Srinivas Neppalli and Tarun Joshi and Michal Malohlava},
      year={2024},
      eprint={2411.16391},
      archivePrefix={arXiv},
      primaryClass={cs.CL},
      url={https://arxiv.org/abs/2411.16391}, 
}

@inproceedings{li2015counterfactual,
    author = {Li, Lihong and Chen, Shunbao and Kleban, Jim and Gupta, Ankur},
    title = {Counterfactual Estimation and Optimization of Click Metrics in Search Engines: A Case Study},
    year = {2015},
    isbn = {9781450334730},
    publisher = {Association for Computing Machinery},
    address = {New York, NY, USA},
    url = {https://doi.org/10.1145/2740908.2742562},
    doi = {10.1145/2740908.2742562},
    booktitle = {Proceedings of the 24th International Conference on World Wide Web},
    pages = {929–934},
    numpages = {6},
    location = {Florence, Italy},
    series = {WWW '15 Companion}
}

@inproceedings{zhu2020joint,
    author = {Zhu, Ziwei and He, Yun and Zhang, Yin and Caverlee, James},
    title = {Unbiased Implicit Recommendation and Propensity Estimation via Combinational Joint Learning},
    year = {2020},
    isbn = {9781450375832},
    publisher = {Association for Computing Machinery},
    address = {New York, NY, USA},
    url = {https://doi.org/10.1145/3383313.3412210},
    doi = {10.1145/3383313.3412210},
    booktitle = {Proceedings of the 14th ACM Conference on Recommender Systems},
    pages = {551–556},
    numpages = {6},
    location = {Virtual Event, Brazil},
    series = {RecSys '20}
}

@ARTICLE{wang2022causal,
    author={Wang, Xiangmeng and Li, Qian and Yu, Dianer and Cui, Peng and Wang, Zhichao and Xu, Guandong},
    journal={IEEE Transactions on Knowledge and Data Engineering}, 
    title={Causal Disentanglement for Semantic-Aware Intent Learning in Recommendation}, 
    year={2023},
    volume={35},
    number={10},
    pages={9836-9849},
    doi={10.1109/TKDE.2022.3159802}
}

@misc{gao2023rag,
      title={Retrieval-Augmented Generation for Large Language Models: A Survey}, 
      author={Yunfan Gao and Yun Xiong and Xinyu Gao and Kangxiang Jia and Jinliu Pan and Yuxi Bi and Yi Dai and Jiawei Sun and Meng Wang and Haofen Wang},
      year={2024},
      eprint={2312.10997},
      archivePrefix={arXiv},
      primaryClass={cs.CL},
      url={https://arxiv.org/abs/2312.10997}, 
}

@inproceedings{saito2020unbiased,
    author = {Saito, Yuta},
    title = {Unbiased Pairwise Learning from Biased Implicit Feedback},
    year = {2020},
    isbn = {9781450380676},
    publisher = {Association for Computing Machinery},
    address = {New York, NY, USA},
    url = {https://doi.org/10.1145/3409256.3409812},
    doi = {10.1145/3409256.3409812},
    booktitle = {Proceedings of the 2020 ACM SIGIR on International Conference on Theory of Information Retrieval},
    pages = {5–12},
    numpages = {8},
    location = {Virtual Event, Norway},
    series = {ICTIR '20}
}

@inproceedings{zhuang2022implicit,
    author = {Zhuang, Shengyao and Li, Hang and Zuccon, Guido},
    title = {Implicit Feedback for Dense Passage Retrieval: A Counterfactual Approach},
    year = {2022},
    isbn = {9781450387323},
    publisher = {Association for Computing Machinery},
    address = {New York, NY, USA},
    url = {https://doi.org/10.1145/3477495.3531994},
    doi = {10.1145/3477495.3531994},
    booktitle = {Proceedings of the 45th International ACM SIGIR Conference on Research and Development in Information Retrieval},
    pages = {18–28},
    numpages = {11},
    location = {Madrid, Spain},
    series = {SIGIR '22}
}

@inproceedings{kentapadi2024grounding,
    author = {Kenthapadi, Krishnaram and Sameki, Mehrnoosh and Taly, Ankur},
    title = {Grounding and Evaluation for Large Language Models: Practical Challenges and Lessons Learned (Survey)},
    year = {2024},
    isbn = {9798400704901},
    publisher = {Association for Computing Machinery},
    address = {New York, NY, USA},
    url = {https://doi.org/10.1145/3637528.3671467},
    doi = {10.1145/3637528.3671467},
    booktitle = {Proceedings of the 30th ACM SIGKDD Conference on Knowledge Discovery and Data Mining},
    pages = {6523–6533},
    numpages = {11},
    location = {Barcelona, Spain},
    series = {KDD '24}
}

@misc{zhou2024trustworthiness,
      title={Trustworthiness in Retrieval-Augmented Generation Systems: A Survey}, 
      author={Yujia Zhou and Yan Liu and Xiaoxi Li and Jiajie Jin and Hongjin Qian and Zheng Liu and Chaozhuo Li and Zhicheng Dou and Tsung-Yi Ho and Philip S. Yu},
      year={2024},
      eprint={2409.10102},
      archivePrefix={arXiv},
      primaryClass={cs.IR},
      url={https://arxiv.org/abs/2409.10102}, 
}

@misc{ni2025trustworthy,
      title={Towards Trustworthy Retrieval Augmented Generation for Large Language Models: A Survey}, 
      author={Bo Ni and Zheyuan Liu and Leyao Wang and Yongjia Lei and Yuying Zhao and Xueqi Cheng and Qingkai Zeng and Luna Dong and Yinglong Xia and Krishnaram Kenthapadi and Ryan Rossi and Franck Dernoncourt and Md Mehrab Tanjim and Nesreen Ahmed and Xiaorui Liu and Wenqi Fan and Erik Blasch and Yu Wang and Meng Jiang and Tyler Derr},
      year={2025},
      eprint={2502.06872},
      archivePrefix={arXiv},
      primaryClass={cs.CL},
      url={https://arxiv.org/abs/2502.06872}, 
}

@misc{ojewale2026audit,
      title={Audit Trails for Accountability in Large Language Models}, 
      author={Victor Ojewale and Harini Suresh and Suresh Venkatasubramanian},
      year={2026},
      eprint={2601.20727},
      archivePrefix={arXiv},
      primaryClass={cs.CY},
      url={https://arxiv.org/abs/2601.20727}, 
}

@inproceedings{vardasbi2020when,
    author = {Vardasbi, Ali and Oosterhuis, Harrie and de Rijke, Maarten},
    title = {When Inverse Propensity Scoring Does Not Work: Affine Corrections for Unbiased Learning to Rank},
    year = {2020},
    isbn = {9781450368599},
    publisher = {Association for Computing Machinery},
    address = {New York, NY, USA},
    url = {https://doi.org/10.1145/3340531.3412031},
    doi = {10.1145/3340531.3412031},
    booktitle = {Proceedings of the 29th ACM International Conference on Information \& Knowledge Management},
    pages = {1475–1484},
    numpages = {10},
    location = {Virtual Event, Ireland},
    series = {CIKM '20}
}

@inproceedings{vardasbi2020cascade,
    author = {Vardasbi, Ali and de Rijke, Maarten and Markov, Ilya},
    title = {Cascade Model-Based Propensity Estimation for Counterfactual Learning to Rank},
    year = {2020},
    isbn = {9781450380164},
    publisher = {Association for Computing Machinery},
    address = {New York, NY, USA},
    url = {https://doi.org/10.1145/3397271.3401299},
    doi = {10.1145/3397271.3401299},
    booktitle = {Proceedings of the 43rd International ACM SIGIR Conference on Research and Development in Information Retrieval},
    pages = {2089–2092},
    numpages = {4},
    location = {Virtual Event, China},
    series = {SIGIR '20}
}

@InProceedings{vardasbi2026adaptive,
    author="Vardasbi, Ali and Penha, Gustavo and Hauff, Claudia and Bouchard, Hugues",
    title="Adaptive Repetition for Mitigating Position Bias in LLM-Based Ranking",
    booktitle="Advances in Bias, Fairness, and Understudied Users in Information Retrieval",
    year="2026",
    publisher="Springer Nature Switzerland",
    address="Cham",
    pages="3--15",
    isbn="978-3-032-12717-4",
    doi = {10.1007/978-3-032-12717-4_1},
    url = {https://link.springer.com/chapter/10.1007/978-3-032-12717-4_1}
}
